\documentclass[acmsmall]{acmart}
\AtBeginDocument{%
  \providecommand\BibTeX{{%
    \normalfont B\kern-0.5em{\scshape i\kern-0.25em b}\kern-0.8em\TeX}}}

\setcopyright{acmcopyright}
\copyrightyear{2020}
\acmYear{2020}
\acmDOI{10.1145/1122445.1122456}

\usepackage{amssymb}

\begin{document}

\title{Deep Residual Neural Networks for Image in Speech Steganography}

\author{Shivam Agarwal}
\authornote{Both authors contributed equally to this research.}
\email{shivamag99@gmail.com}
\author{Siddarth Venkatraman}
\authornotemark[1]
\email{siddarthegreat@gmail.com}
\affiliation{%
  \institution{Manipal Institute of Technology}
  \city{Manipal}
  \state{Karnataka}
}

\pdfoutput=1

\begin{abstract}
Steganography is the art of hiding a secret message inside a publicly visible carrier message. Ideally,  it is done without modifying the carrier, and with minimal loss of information in the secret message. Recently, various deep learning based approaches to steganography have been applied to different message types. We propose a deep learning based technique to hide a source RGB image message inside finite length speech segments without perceptual loss. To achieve this, we train three neural networks; an encoding network to hide the message in the carrier, a decoding network to reconstruct the message from the carrier and an additional image enhancer network to further improve the reconstructed message. We also discuss future improvements to the algorithm proposed.
\end{abstract}




\settopmatter{printacmref=false}
\setcopyright{none}
\renewcommand\footnotetextcopyrightpermission[1]{}
\pagestyle{plain}
\fancyfoot{}
\maketitle
\thispagestyle{empty}

\section{Introduction}
In Steganography, the main goal is secret communication. The sender encodes an image in an audio sample such that the receiver can decode the secret message but, an uninformed listener should not be able to detect the presence of the hidden image. 

\noindent
We introduce a method for image-in-speech steganography using a trainable end-to-end neural network. An encoder network receives a cover speech segment and a secret image and outputs an embedded speech segment. The decoder network attempts to reconstruct the hidden image from the embedded speech segment. A separate network is used to enhance the reconstructed image quality. 

\noindent
Classical steganography methods usually use heuristics to decide how much to modify each pixel. For example, some algorithms change the least significant bits of some selected parts of the cover[20]. Whereas, some change the mid-frequency  components in the frequency domain [21]. These heuristics are usually not robust to variation in the environment. While, Deep Neural Networks are trained to achieve the objective of interest thereby providing much better steganography quality. 

\noindent
Recently, Deep Learning methods have been successfully applied to image-in-image steganography[1] and audio-in-audio steganography[2]. In this work we present a method for image-in-audio steganography using deep residual neural networks for encoding, decoding and enhancing the secret image. 
\section{Related Work}
A large number of steganography methods have been proposed over the years. Most of these are applied on Images[15] or Audio[19]. The most common approach is to manipulate the Least Significant Bits of the cover to embed the secret information. This can be done by simply replacing the LSBs of the cover uniformly or adaptively[16] with the secret data. Simple statistical analysis of the embedded carrier reveals presence of a hidden messages. More complex approaches such as HUGO[17] provide better steganography by preserving the input statistics. Another example is WOW (Wavelet Obtained Weights)[18] that penalizes distortion of predictable regions of the input data using a bank of directional filters.

\noindent
Recently, neural networks have been applied to Steganography and Steganalysis. Neural networks provide much better results compared to the traditional techniques because of their ability to model the entire hiding and revealing process. Additionally, neural nets can be trained end-end further reducing cumulative errors caused while using individual modules. S.Baluja[1] proposed an encoder-decoder architecture for image-in-image steganography using end-to-end neural networks. A similar technique was proposed by F. Kruek et al[2] for audio-in-audio steganography. However, these techniques have been applied to same message types, either image-in-image or audio-in-audio only. In this work we present a method to hide images in finite length speech segments.

\section{Problem Formulation}
In this section we define the notation we contrive for the task of image in audio steganography and define the notation for the rest of the paper. We denote the set of fixed length acoustic feature vectors \(\textbf{X}\in\mathbb{R}^{m\times n}\)
and the set of all fixed resolution images as \(\textbf{Y }\in \mathbb{R}^{m \times n}\). 
The goal of this model is to conceal an image message in a fixed length speech segment. The steganography model gets the input carrier spectrogram denoted by \textbf{C} such that $\textbf{C} \in \mathbf{X}$ and the input message as \textbf{M} such that all $\textbf{M} \in \mathbf{Y}$. The model outputs the embedded carrier spectrogram denoted as $\textbf{\^C} \in \mathbf{X}$ and the reconstructed image denoted as $\textbf{\^M}\in \mathbf{Y}$. \textbf{\^C} and \textbf{\^M} should satisfy the following conditions:
\begin{enumerate}
\item 
\textbf{\^C} and \textbf{\^M} should be similar to \textbf{C} and and \textbf{M} respectively.  

\item \textbf{\^M} should be reconstructible from \textbf{\^C} without any perceptual loss.
\item
An uninformed human listener must not be able to detect the presence of a hidden message in \textbf{\^C}. 
\end{enumerate}

\section{Model Architecture}
In this section we present the the model architecture for the problem in hand. Our model consists of two parts:
\subsection{Main Model}
The main architecture is composed of 2 blocks; the Hiding Block(H), the Reveal Block(R). The Hiding Block takes in both \textbf{C} (input speech segment) and \textbf{M} (the input image). Precisely, the \textbf{C} is carrier spectrogram and \textbf{M} is a single channel of the image to be hidden. \textbf{C} is passed through stack of gated convolutions [5] and \textbf{M} is passed through a stack of vanilla convolutional layers with a receptive field of 3x3. The convolution stride is fixed to 1 pixel. The two outputs are concatenated and passed through downsampling encoder followed by a upsampling decoder. There are also residual connections between the downsampling and upsampling block at the same scales. This block outputs \textbf{\^C} (the embedded carrier). 

\noindent
\textbf{\^C} (the embedded carrier spectrogram) generated by H (Hiding block) is passed as an input to R (Reveal block). We use an encoder-decoder network similar to the Hidding Block, H. In this block the reconstructed message, \textbf{M} is obtained from the embedded carrier, \textbf{\^C}. All layers in the reveal block is attached with ReLU activation[10] and sigmoid activation[12] function for for the last layer. Lastly, the appropriate parameters are found by minimizing the Mean Squared Error(MSE) between the carrier spectrogram and the embedded carrier spectrogram and between the input image and the reconstructed image.
\begin{figure*}[h!]
\centering
\includegraphics[width=1\textwidth, height=0.4\textwidth, keepaspectratio]{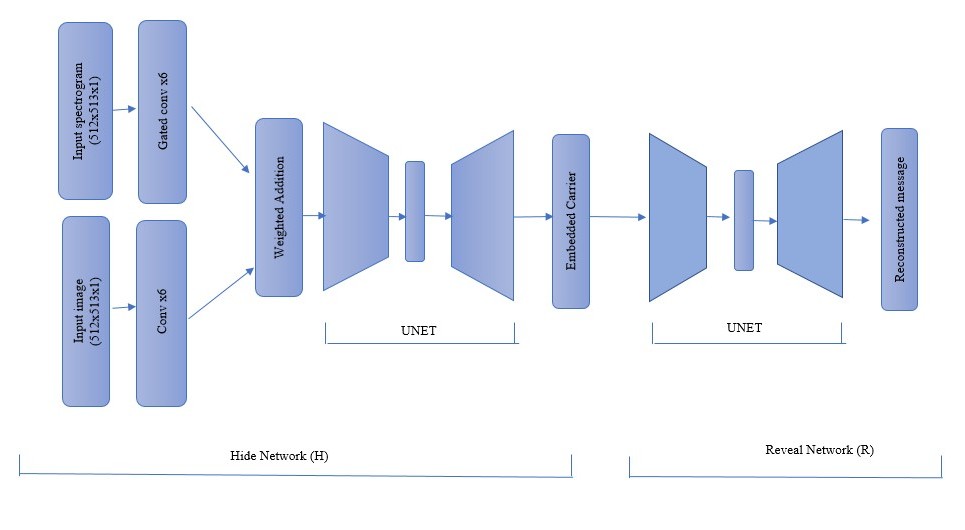}
\caption{Model Architecture}
\end{figure*} 
In addition, we add a small component of the Structured Similarity(SSIM)[7]. SSIM is a useful metric for evaluating the quality of the predictions.The SSIM for two signals \(x\) and \(y\) is defined as:
\[\mathbf{SSIM=\frac{(2\mu_x\mu_y+c_1)(2\sigma_{xy}+c_2)}{(\mu_x^2+\mu_y^2+c_1)(\sigma_x+\sigma_y+c_2)}}\]
\noindent
Where \(\mu\) and \(\sigma\) are the the local mean and variance of \(x\) and \(y\). \(c_1=(k_1L^2)\) and \(c_2=(k_2L^2)\) are two sizable variables to stabilize the division with weak denominator and \(L\) is the dynamic range of signal values. We use the default values of \(k_1=0.01\) and \(k_2=0.03\) as in the original paper[7]. Since SSIM is upperbounded to 1 we need to maximise it. Instead we minimise:
\[\mathbf{L_{SSIM}(x,y)=[1-SSIM(x,y)]}\]
\noindent
The final loss function is given by the equations:
\[\mathbf{L_{C}=\lambda_{C_1}(C-H(C,M))^2 + \lambda_{C_2}(L_{SSIM}(C,\hat{C})}\]

\[\mathbf{L_{M}=\lambda_{M_1}(M-R(\hat{C}))^2 + \lambda_{M_2}(L_{SSIM}(M,\hat{M})}\]

\[\mathbf{L=\lambda_{M}L_M+\lambda_{C}L_C}\]
\subsection{Message Enhancer}
We also trained a fully convolutional image enhancer network. This module removed small noises from the reconstructed images on the receiver side. We use a U-net architecture with skip connections from the encoder and decoder blocks at the same scale with downsampling while encoding and upsampling while decoding and ReLU activation function.

\section{Experimental Setup}

In this section, we present our experimental results. First, we describe our experimental setup. Then we evaluate our model. Lastly, we show visual analysis of the images and spectrograms. We have implemented our code with tensorflow [3]. Codes for this experiment can be found \href{https://github.com/shivamag125/image_speech_steganography}{\underline{here.}}
\begin{figure*}[h!]
\centering
\includegraphics[width=1\textwidth, height=0.5\textwidth]{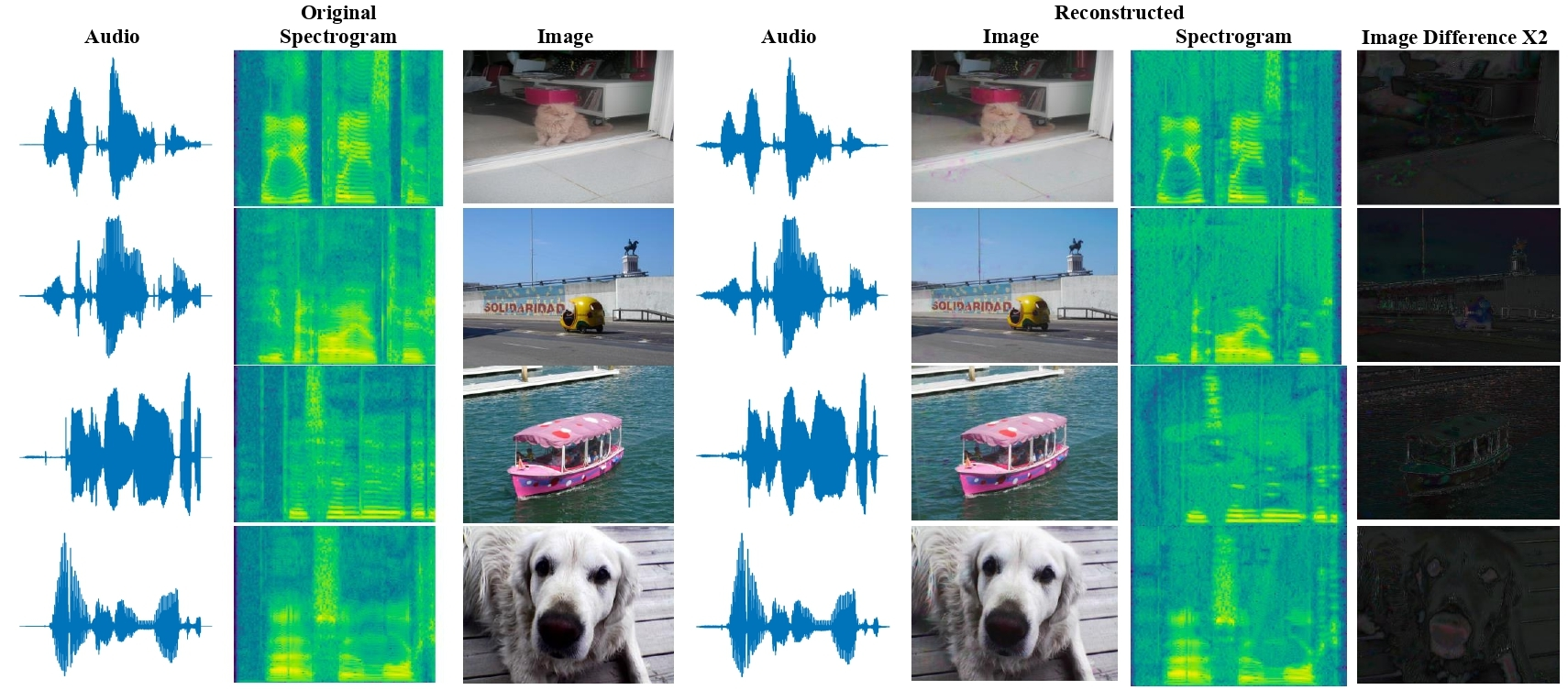}
\label{results}
\caption{Results}
\end{figure*}

\subsection{Experimental setup:}

We evaluated our approach on TIMIT[6] for audio segments and Pascal Voc dataset for images. Each audio sample was sampled at 16 kHz and represented as a power spectrum by taking the Short Term Fourier Transform(STFT) according to the parameters shown in Table 2. These parameters were obtained to achieve an STFT of size 512*513 and whose inverse can be obtained. The magnitude of STFT is passed as input to the model. The model produces the magnitude spectrogram of the embedded carrier. We use the same phase for estimating the time domain audio sample by calculating the Inverse STFT. Each image was resized to 512*513 from the Pascal Voc dateset. The model was trained using the Adam optimizer[4] for 203K steps for TIMIT. The losses were balanced according to Table 1. 

\noindent
The model is trained with single greyscale images. For color images, individual Red, Green and Blue channels are separately passed to the model and the corresponding audio samples are concatenated to form the carrier. RGB images are generated with this mechanism from the message decoder network. These images are used to train the image enhancer network that removes small noises in the reconstructed images on the receiver side. 

\begin{table}[h!]
\centering
\caption{Loss balance parameters}
\label{Loss Balance}
\begin{tabular}{|l|l|l|l|l|l|}
\hline
\(\lambda_{C_1}\) & \(\lambda_{C_2}\) & \(\lambda_{M_1}\) & \(\lambda_{M_2}\) & \(\lambda_{C}\) & \(\lambda_{M}\) \\ \hline
0.995         & 0.005         & 0.85          & 0.15          & 0.69        & 1           \\ \hline
\end{tabular}
\end{table}

\begin{table}[h!]
\caption{STFT Parameters}
\centering
\begin{tabular}{|l|l|l|l|l}
\cline{1-4}
Duration of audio & FFT Length & Frame Step & Frame Length &  \\ \cline{1-4}
1 second          & 1024       & 31         & 1024        &  \\ \cline{1-4}
\end{tabular}
\end{table}

\begin{figure}[h!]
\centering
\includegraphics[width=0.4\textwidth]{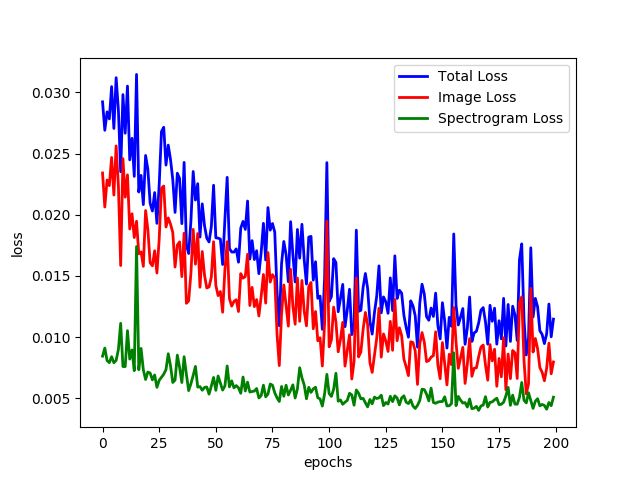}
\caption{Training curve}
\end{figure}

\subsection{Analysis:}

To evaluate the results obtained we compute the Mean Absolute Error:
\[\mathbf{MAE}=\frac{1}{N}\sum_{i=1}^N\mod(x_i-x_i')\]
Where \(x_i\) and the \(x_i'\) are the true and the predicted values respectively. For the carrier we evaluate the Pearson correlation coefficient. 
\[\mathbf{r=\frac{n(\sum xy)-(\sum x)(\sum y)}{\sqrt{[n\sum x^2-(\sum x)^2][n\sum y^2-(\sum y)^2]}}}\]
Where x and y are the two sets of data and n is the number of samples. We also evaluate the MAE for each audio sample. Experimental results are tabulated in table III

\begin{table}[h!]
\caption{Results obtained }
\label{tab:my-table}
\begin{tabular}{|l|l|l|l|}
\hline
\textbf{Parameter}                                                      & \textbf{\begin{tabular}[c]{@{}l@{}}Model+SSIM+\\ Image Enhancer\end{tabular}} & \textbf{Model + SSIM} & \textbf{Base Model} \\ \hline
MAE(Image)                                                                 & 8.1673                                                                       & 7.3529                & 13.3901             \\ \hline
SSIM (Image)                                                               & 0.9486                                                                       & 0.9246                & 0.8666              \\ \hline
\begin{tabular}[c]{@{}l@{}}MAE\\ (Spectrogram)\end{tabular}                & 0.0715                                                                       & 0.0715                & 0.0633              \\ \hline
\begin{tabular}[c]{@{}l@{}}SSIM\\ (Spectrogram)\end{tabular}               & 0.99878                                                                      & 0.99878               & 0.9987              \\ \hline
\begin{tabular}[c]{@{}l@{}}Pearson Correlation \\ Coefficient\end{tabular} & 0.4943                                                                       & 0.4943                & 0.4999              \\ \hline
\end{tabular}
\end{table}

\section{Discussions and Future Work}
We have shown that deep learning steganography techniques using encoder-decoder setup can be used for steganography where the carrier and source message data types are different. More sophisticated methods of hyperparameter search can probably be used to further improve the results shown above. Future work could also involve better techniques to train networks with losses that compete with one another.

\noindent
The output images were best when coupled with the image enhancer. This ends up slightly altering the color palette of some images, even though the outputs would fool most people when judged on their own.  In the future, we could experiment with more sophisticated autoregressive decoders which might eliminate the need for a denoiser. We can also boost the audio quality by using vocoder architectures to account for losses while converting from spectrogram to time domain audio. Finally, hiding multiple images in short audio clips can definitely be studied in the future to increase its effectiveness. 

\section{Acknowledgment}
We thank Project MANAS for supporting us with the necessary resources.

\end{document}